\UseRawInputEncoding
\documentclass[aps,groupedaddress,superscriptaddress,preprint]{revtex4}

\usepackage{graphicx}
\usepackage{color}
\usepackage{amsmath}
\usepackage{dcolumn}
\usepackage{bm}
\usepackage{multirow}
\usepackage{epstopdf}

\begin{document}

\title{$^{181}$Ta Nuclear quadrupole resonance study \\of non-centrosymmetric superconductor PbTaSe$_2$}

\author{K. Yokoi$^*$}
\affiliation{Department of Physics, Osaka University, Toyonaka, Osaka 560-0043, Japan}

\author{M. Yashima}
\affiliation{Graduate School of Engineering Science, Osaka University, Osaka 560-8531, Japan}

\author{H. Murakawa$^*$}
\affiliation{Department of Physics, Osaka University, Toyonaka, Osaka 560-0043, Japan}

\author{H. Mukuda$^*$}
\affiliation{Graduate School of Engineering Science, Osaka University, Osaka 560-8531, Japan}

\author{K. Yamauchi}
\affiliation{ISIR-SANKEN, Osaka University, Osaka 567-0047, Japan}

\author{T. Oguchi}
\affiliation{ISIR-SANKEN, Osaka University, Osaka 567-0047, Japan}

 \author{H. Sakai}
\affiliation{Department of Physics, Osaka University, Toyonaka, Osaka 560-0043, Japan}
\affiliation{PRESTO, Japan Science and Technology Agency, Kawaguchi, Saitama 332-0012, Japan}

 \author{N. Hanasaki}
\affiliation{Department of Physics, Osaka University, Toyonaka, Osaka 560-0043, Japan}

\begin{abstract}
{
We report on a pure $^{181}$Ta-nuclear quadrupole resonance (NQR) measurement of PbTaSe$_2$ at zero magnetic field, which has the advantage of directly probing the intrinsic superconducting phase and electronic states of the TaSe$_2$ layer. We observed the $^{181}$Ta-NQR spectrum of the intrinsic structure with space group $P$\={6}$m2$, which agrees well with density functional theory (DFT) calculations. The nuclear spin relaxation rate ($1/T_1$) shows an exponential decrease well below $T_{\rm_c}$, indicating that the superconducting state is fully gapped in the framework of Bardeen-Cooper-Schrieffer  (BCS) theory. The gap size obtained by $^{181}$Ta-NQR was smaller than the value in previous reports, which may imply that the Fermi surfaces composed of Ta-5$d$ orbitals, where the average pairing interactions are expected to be weaker than in BCS model, are primarily probed. The temperature dependence of $1/T_1$ below $T_{\rm_c}$ can be reproduced well by the superposition of quadrupole and magnetic relaxation mechanisms, together with the distribution of superconducting gap size inherent to multiple Fermi surfaces theoretically proposed in PbTaSe$_2$.
}
\end{abstract}

\maketitle

\section{Introduction}
Space inversion symmetry breaking in crystal structure makes asymmetric spin-orbit coupling (ASOC), which often leads to unique superconducting properties beyond the conventional Bardeen-Cooper-Schrieffer (BCS) framework\cite{1,2}.
For example, unconventional features have been experimentally reported in CePt$_3$Si\cite{3}, CeMSi$_3$ (M = Rh, Ir)\cite{4,5,6,7}, and Li$_2$(Pd, Pt)$_3$B\cite{8,9,10,11}, R$_2$C$_3$ (R = La, Y)\cite{12,13,14}.

The transition metal dichalcogenide-based layer compound PbTaSe$_2$ without inversion symmetry is a type-II superconductor with a superconducting transition temperature of $T_{\rm_c} \sim 3.8$ K\cite{15}. 
As shown in Fig. 1(a), the crystal structure of PbTaSe$_2$ is composed of 2H-TaSe$_2$ and intercalated Pb layers\cite{16}. 
The strong ASOC stemming from heavy elements (Pb and Ta) lifts the spin degeneracy and can induce a parity-mixed superconducting state\cite{17,18}. 
Furthermore, PbTaSe$_2$ possesses topological electronic states such as nodal line fermions and drumhead surface states\cite{19,20}, which have the potential to realize Dirac/Weyl superconducting states\cite{21} and topological superconductivity in the surface\cite{22}.

Many experimental studies have been carried out for detection of the unique superconducting state of PbTaSe$_2$.
However, so far only the signatures of conventional BCS-type $s$-wave superconductivity have been reported, including heat capacity\cite{23,24,25}, thermal conductivity\cite{26}, tunnel diode oscillator\cite{27}, STM\cite{28}, and $\mu$SR\cite{29}. 
Microscopic evidence for conventional BCS-type $s$-wave superconductivity was also reported by $^{207}$Pb nuclear magnetic resonance (NMR) measurements\cite{30}, which were performed for vortex states in an external magnetic field  $\mu_{0}H = 0.19$ T close to  $\mu_{0} H_{\rm_{c}2}^{c} (T=0) \approx 0.3$ T\cite{23,24,25,26}. 
To elucidate hidden electronic states and superconducting properties in zero magnetic field, it is also desirable to perform $^{181}$Ta NQR, which allows local electronic states of TaSe$_2$ layer to be probed directly. 
The band structure of PbTaSe$_2$ is characterized by multiple Fermi surfaces (FSs) as shown in Fig. 1(b) structured around the $\Gamma$ point in $k$-space dominated by Se-4$p$ and Ta-5$d$ orbitals and around K and H points dominated by Ta-5$d$ and Pb-4$p$ derived orbitals\cite{31}. 
A multiple gap scenario related to multiband superconductivity in PbTaSe$_2$ has been proposed by some experiments and theory\cite{23,25,26,27,28,29,30,31}. 
Taking the possible difference in hyperfine coupling with the nucleus for each FS into account, it is advantageous to select the Ta site for comparison with Pb site results to aid further understanding of multiband superconductivity of PbTaSe$_2$.

In this paper, we report a {\it pure} NQR study of the $^{181}$Ta-nucleus in PbTaSe$_2$ at zero fields, expected to be a sensitive probe of local electronic bands of the multiple FSs of TaSe$_2$ layer. 
In our results we found two inequivalent Ta sites: the Ta(1) site is identified as an intrinsic site derived from a non-centrosymmetric structure for $P$\={6}$m2$ symmetry using $^{181}$Ta-NQR spectrum analysis combined with density functional theory (DFT) calculations, while the Ta(2) site is expected to arise from the inevitable structural impurity. 
The nuclear spin-lattice relaxation rate ($1/T_1$) at the intrinsic Ta(1) site exhibits an exponential decrease well below $T_{\rm_c}$, which indicates that the superconducting state is fully gapped in PbTaSe$_2$. 
Temperature dependence of $1/T_1$ is reproduced by the superposition of {\it quadrupole} and {\it magnetic} relaxation mechanisms in addition to the existence of distribution of gap size. 
The gap parameter $2\Delta/k_{\rm_B}T_{\rm_c}$ obtained in this work was near 3.1, which is smaller than the value obtained in other experiments. 
This may suggest that FSs dominantly composed of Ta-5$d$ orbitals can be primarily probed by $^{181}$Ta-NQR.

\begin{figure}[t]
	\begin{center}
		\includegraphics[width=0.8\linewidth]{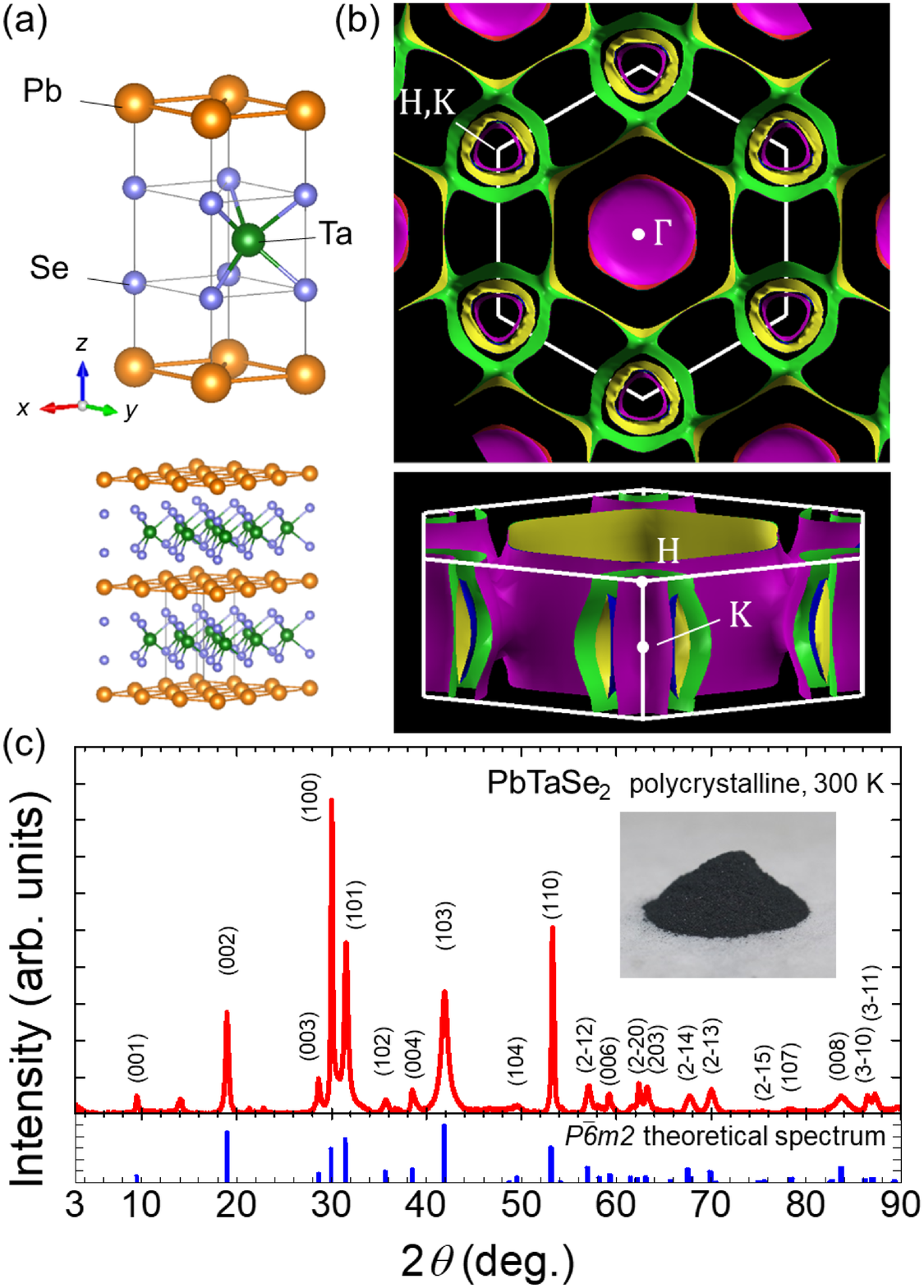}
		\caption[Structure]{\label{fig1}(Color online) (a) Crystal structure of PbTaSe$_2$ with space group $P$\={6}$m2$. (b) Fermi surfaces (FS) obtained by the DFT calculation. FSs at the $\Gamma$ point and an outer FS around K, H points are mainly composed of Ta-5$d$, while an inner cylindrical FS around K, H points are mainly composed of Pb-4$p$ orbitals. (c) Powder X-ray diffraction spectrum of PbTaSe$_2$ at room temperature is shown in the upper panel, which is consistent with the diffraction pattern simulation for space group $P$\={6}$m2$ in the lower panel.
		 }
	\end{center}
\end{figure}
 
\section{Experiment and Calculation}
Polycrystalline samples of PbTaSe$_2$ were synthesized via a solid-state reaction.
Stoichiometric amounts of Pb (powder, 99.9\%), Ta (powder, 99.9\%), and Se (grain, 99.999\%) were placed in a silica tube, sealed under vacuum, and heated for $3 - 5$ days\cite{15,16}. 
The dc and ac susceptibilities were measured by a magnetic property measurement system (MPMS, Quantum Design) and {\it in situ} NQR coil, respectively. 
The NQR measurement was performed using a conventional phase-coherent-type spectrometer. 
The $^{181}$Ta-NQR spectrum was obtained by sweeping the frequency and integrating the spin-echo intensity.
The nuclear spin-lattice relaxation rate ($1/T_1$) was measured by the saturation-recovery method in a temperature range of $1.6-20$ K. 
The observed nuclear magnetization recovery curve was well-fitted by $1 - M(t)/M(\infty) = 1/42\exp(-3 t/T_{1}) + 18/77\exp(-10t/T_{1}) + 49/66\exp(-21t/T_{1})$\cite{32}. 
To evaluate the NQR frequency theoretically, the electronic field gradient at the Ta site was calculated using the all-electron full-potential linearized-augmented-plane-wave program package HiLAPW\cite{33}. 
Additionally, the FSs were calculated using the WIEN2k package\cite{34} and drawn by Xcrysden. 
These calculations were performed using GGA-PBE exchange-correlation potential\cite{35} and the full-potential LAPW (FLAPW) basis set taking into account the spin-orbit interaction. 
The lattice parameters and atomic positions were fixed at their experimental values to obtain the band structure.

\section{Results and Discussion}
Figure 1(c) shows the result of the powder X-ray diffraction measurement, which is consistent with the space group $P$\={6}$m2$. 
The lattice parameters were estimated to be $a = 3.4445(5) $\AA\ and $c = 9.3798(18)$\AA\ by the Rietveld analysis. 
They are close to the values reported previously\cite{16,30,36}. 
Although impurity diffraction peaks that corresponded to the 2H-TaSe$_2$ were observed, 
the results from two-phase Rietveld analysis show that the volume fraction of TaSe$_2$ is less than 5\%, meaning the small amount of impurity phase has no influence on the NQR result.


\begin{figure}[t]
	\begin{center}
		\includegraphics[width=0.9\linewidth]{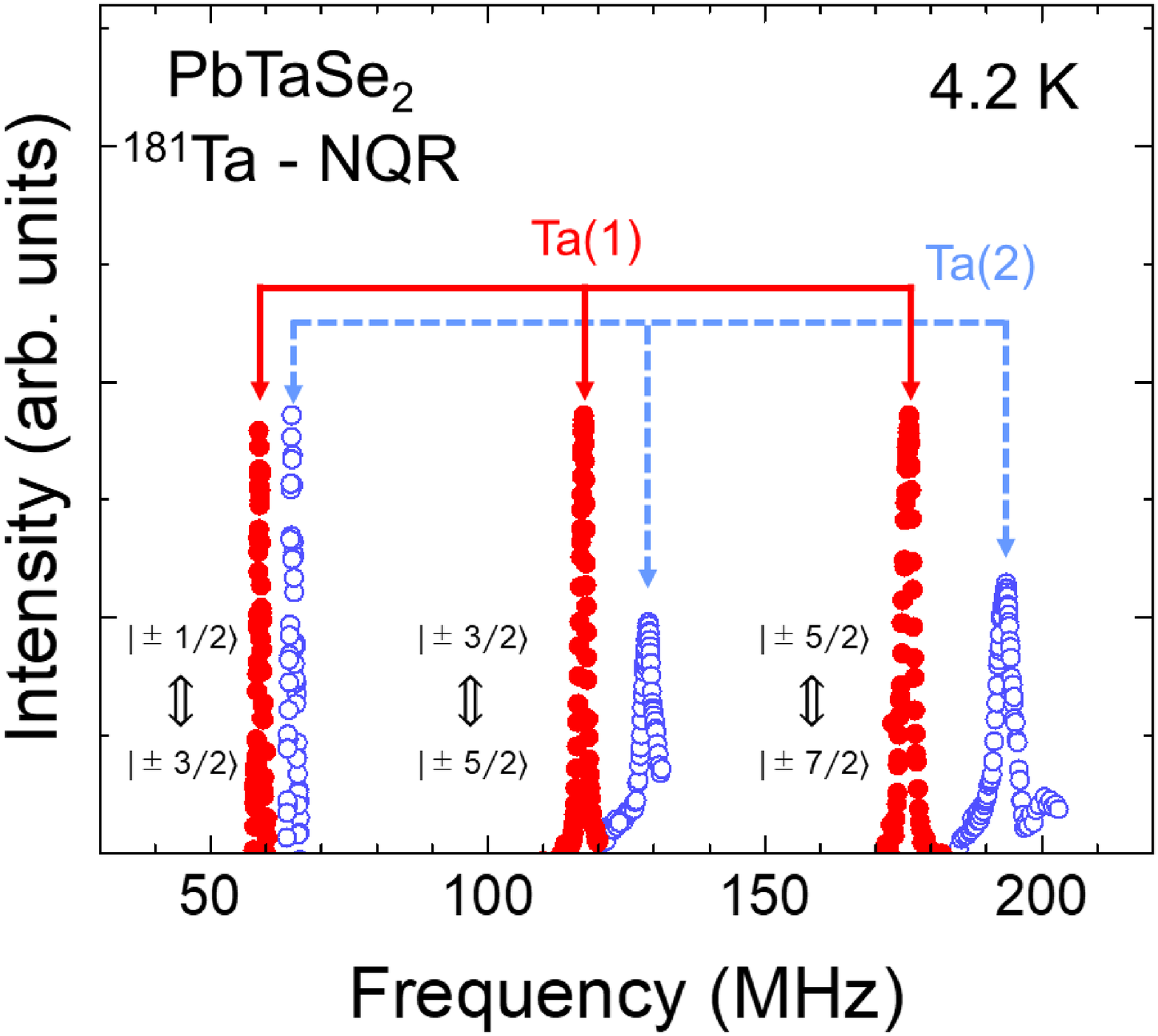} 
		\caption[Structure]{\label{spectrum}(Color online) 
			$^{181}$Ta-NQR spectrum of PbTaSe$_2$ at 4.2 K. Three resonance lines denoted as Ta(1) are reproduced by $\nu_{\rm_Q}()1) =58.7 \pm 0.5$ MHz and $\eta(1)=0.00\sim0.02$, corresponding to the intrinsic Ta site for PbTaSe$_2$ with $P$\={6}$m2$ symmetry. The other broader ones denoted as Ta(2) are reproduced by $\nu_{\rm_Q} (2)=64.5 \pm 0.5$ MHz and $\eta(1)=0.03\sim0.04$, corresponding to the extrinsic Ta site slightly affected by in-plane disorder.
			}
	\end{center}
\end{figure} 

Figure 2 shows the $^{181}$Ta-NQR spectrum of PbTaSe$_2$ at $T = 4.2$ K. Six resonance peaks were observed at $f \approx 58.7, 64.8, 117.3, 128.9, 176.1$, and $193.5 \pm 0.2$ MHz.
In general, nuclear quadrupole interaction is described by the following Hamiltonian,
\begin{equation}
{\cal H}_{Q} = \frac{e^{2} q Q}{4I(2I-1)} \{3 I_{z}^{2} -I(I+1) +\eta(I_{x}^{2}-I_{y}^{2})\},
\end{equation}
where $eQ$ is the nuclear quadrupole moment, $eq = V_{zz}$ is the electronic field gradient (EFG) along the principal axis defined by the maximum EFG direction, and $\eta = |V_{\rm_{xx}}-V_{\rm_{yy}}|/V_{\rm_{zz}}$ is an asymmetric parameter of the EFG. 
Here the NQR frequency is defined as $\nu_{\rm_Q} = (3e^{2}qQ)/2I(2I-1)h$.
In the case of $^{181}$Ta (nuclear spin $I = 7/2$), the energy level of the nuclear magnetic moment is split into four levels ($m = \pm1/2, \pm3/2, \pm5/2, \pm7/2$) by the nuclear quadrupole interaction, and thus, three resonance peaks per Ta site should be observed in the $^{181}$Ta-NQR spectrum. 
Therefore, the observation of the six NQR peaks indicated the presence of two inequivalent Ta sites in the sample, where a single Ta site per unit cell is expected for ideal crystal structure with space group $P$\={6}$m2$ in PbTaSe$_2$. 
Three narrower peaks at $58.7, 117.3$, and $176.1 \pm 0.2$ MHz are reproduced for the case of $\nu_{\rm_{Q}}(1) = 58.7\pm0.5$ MHz and $\eta(1) = 0.00\sim0.02$, which is denoted as the Ta(1) site, while the other three peaks with the broad tails at $64.8, 128.9$, and $193.5 \pm 0.2$ MHz can be reproduced for the case of $\nu_{\rm_{Q}}(2)=64.5\pm0.5$ MHz and $\eta(2)=0.03\sim~0.04$, which is denoted as the Ta(2) site. 
The $\eta(1)\sim0$ represents the axial symmetry around the Ta atom, indicating Ta(1) is an intrinsic Ta site with $P$\={6}$m2$ structure because the Ta site is located on the $C_3$ rotation axis along the $c$ axis in $P$\={6}$m2$.
In this context, the Ta(2) site with a finite value of $\eta(2)$ should be assigned to the impurity site, which arises from the impurity phase that was not observed in X-ray spectrum. 
Note that the comparable intensity of the Ta(2) signal to Ta(1) ensures that this impurity site does not arise from TaSe$_2$, the amount of which is only 5\%.


\begin{table}[t]
	\begin{center}
		\caption[Structure]{Experimental and theoretical values of $\nu_{\rm_Q}$ and $\eta$. Experimental values of $\nu_{\rm_Q}$ for two inequivalent Ta(1) and Ta(2) sites were determined by Lorentzian fitting, and their error bars defined as the FWHM. Values of $\eta$ were determined to reproduce the experimental resonance frequencies.
		\label{table2}}
		\begin{tabular}{ccc}
			&\qquad$\nu_{\rm_Q}$ (MHz) &\qquad$\eta$ \\
			\hline\hline
			Experiment   &  &\\
			Ta(1) site &  \qquad$58.7\pm0.5$ & \qquad$0.00\sim0.02$\\
			Ta(2) site &  \qquad$64.5\pm0.5$ & \qquad$0.03\sim0.04$\\
			\hline
			DFT calculation &   &\\
			$P$\={6}$m2$&\qquad57.78&\qquad0.00\\
			$P6_{3}/mmc$&\qquad46.39&\qquad0.00\\
		\end{tabular}
	\end{center}
\end{table} 

For further verification, the values of $\nu_{\rm_Q}$ and $\eta$ for Ta sites with $P$\={6}$m2$ symmetry were simulated by DFT calculations. 
The calculated values are $\nu_{\rm_Q}=57.8$ MHz and $\eta = 0.0$ as shown in Table I, which is very close to the experimental values assigned to the Ta(1) site.
Consequently, we determined that the Ta(1) site was derived from the non-centrosymmetric structure of PbTaSe$_2$ with $P$\={6}$m2$ symmetry. 
The calculation of $P6_{3}/mmc$, which is another possible structure in intercalated transition metal dichalcogenides\cite{16}, does not agree with the experimentally obtained $\nu_{\rm_Q}(2)$ for the Ta(2) site, as shown in Table I.
The $1/T_1$ measurement at Ta(2) shows similar temperature dependence and absolute value to those of Ta(1), indicating that the electronic state and superconductivity in the impurity phase are close to those in the intrinsic one. 
Pb defects are verified by scanning electron microscope-energy dispersive X-ray spectroscopy (SEM-EDS) at approximately 5\% in our sample. 
Those defects can change the local electric field, and break the in-plane symmetry to result in a finite $\eta$.
Since one Pb defect may affect the six nearest neighboring Ta sites, an amount of $4 - 6\%$ defects can explain the comparable NQR intensity of Ta(1) and Ta(2) sites.
Thus, we assume that Ta(2) comes from the Ta site around Pb defects in the intrinsic $P$\={6}$m2$ structure.

\begin{figure}[b]
	\begin{center}
		\includegraphics[width=0.9\linewidth]{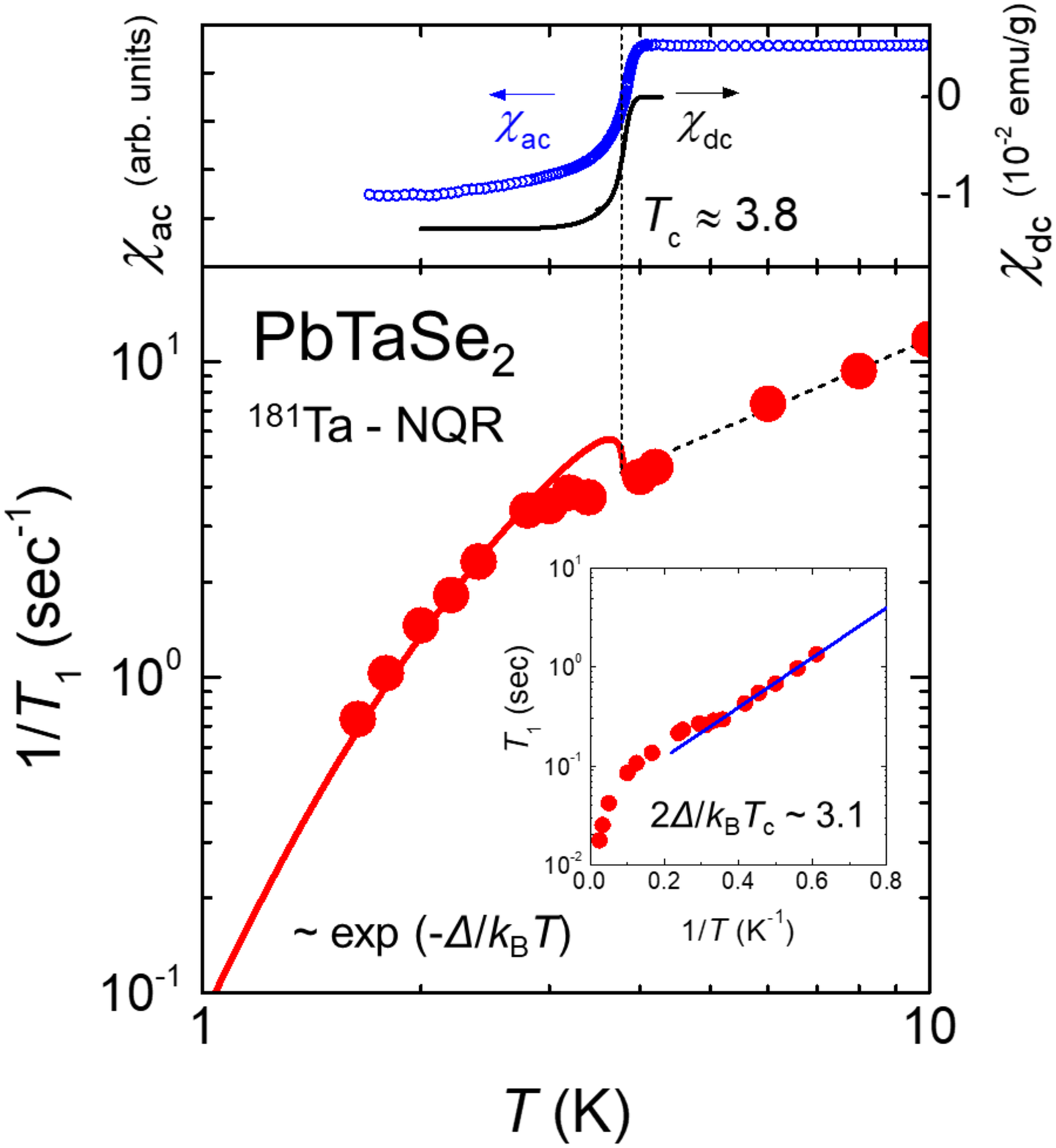} 
		\caption[Structure]{\label{T1}(Color online) 
			$T$ dependence of $1/T_1$ at the Ta(1) sites in PbTaSe$_2$. The upper panel shows the $T$ dependence of AC and DC susceptibilities, which proves $T_c \sim 3.8$ K. Above $T_c$, $1/T_1$ is proportional to $T$, indicating that PbTaSe$_2$ is a non-correlated metal. $1/T_1$ shows exponential decrease below $T_c$, which is evidence for fully-gapped superconductivity. The superconducting gap size $2\Delta$ obtained by the Arrhenius plot of $T_1$ is $3.09 \pm 0.07 k_{\rm_B}T_{\rm_c}$ (inset).}
	\end{center}
\end{figure} 

Next, we focus on the nuclear spin-lattice relaxation rate, $1/T_1$. 
Figure 3 shows the temperature dependence of $1/T_1$ for the Ta(1) site, which was measured from the NQR peaks corresponding to $m=\pm1/2 \leftrightarrow\pm3/2$ transition.
The upper panel of this figure shows the $T$ dependence of the AC susceptibility measured using an {\it in situ} NQR coil (circles), together with that of the DC susceptibility by MPMS (solid line), indicating that bulk superconductivity takes place below $T_{\rm_{c}} \sim 3.8$ K, which is consistent with the values in previous reports\cite{15,19,20,23,24,25,26,27,28,29,30,36}. 
At normal states ($T > T_{\rm_c}$), $1/T_1$ is proportional to temperature, which is generally seen in non-correlated normal metals. 
Note that $^{181}$Ta-NQR is a good probe to reveal charge fluctuations through coupling with the local electric field gradient at the Ta site. 
In the case of Ta$_3$Pd$_4$Te$_{16}$, the $^{181}$Ta-NQR study revealed a large enhancement in $1/T_{1}T$ due to the charge fluctuations derived from charge density wave (CDW) instability\cite{37}. 
As for the case of PbTaSe$_2$, the $T$ dependence of $1/T_1$ without anomalies above $T_{\rm_c}$ suggests that the charge fluctuations are negligible, although the parent layered compound TaSe$_2$ shows CDW order\cite{38,39}, and Pb$_x$TaSe$_2$ ($x = 0.25 - 0.75$) was expected to show charge fluctuations from the Raman spectroscopy\cite{40}.

\begin{figure}[t]
	\begin{center}
		\includegraphics[width=0.85\linewidth]{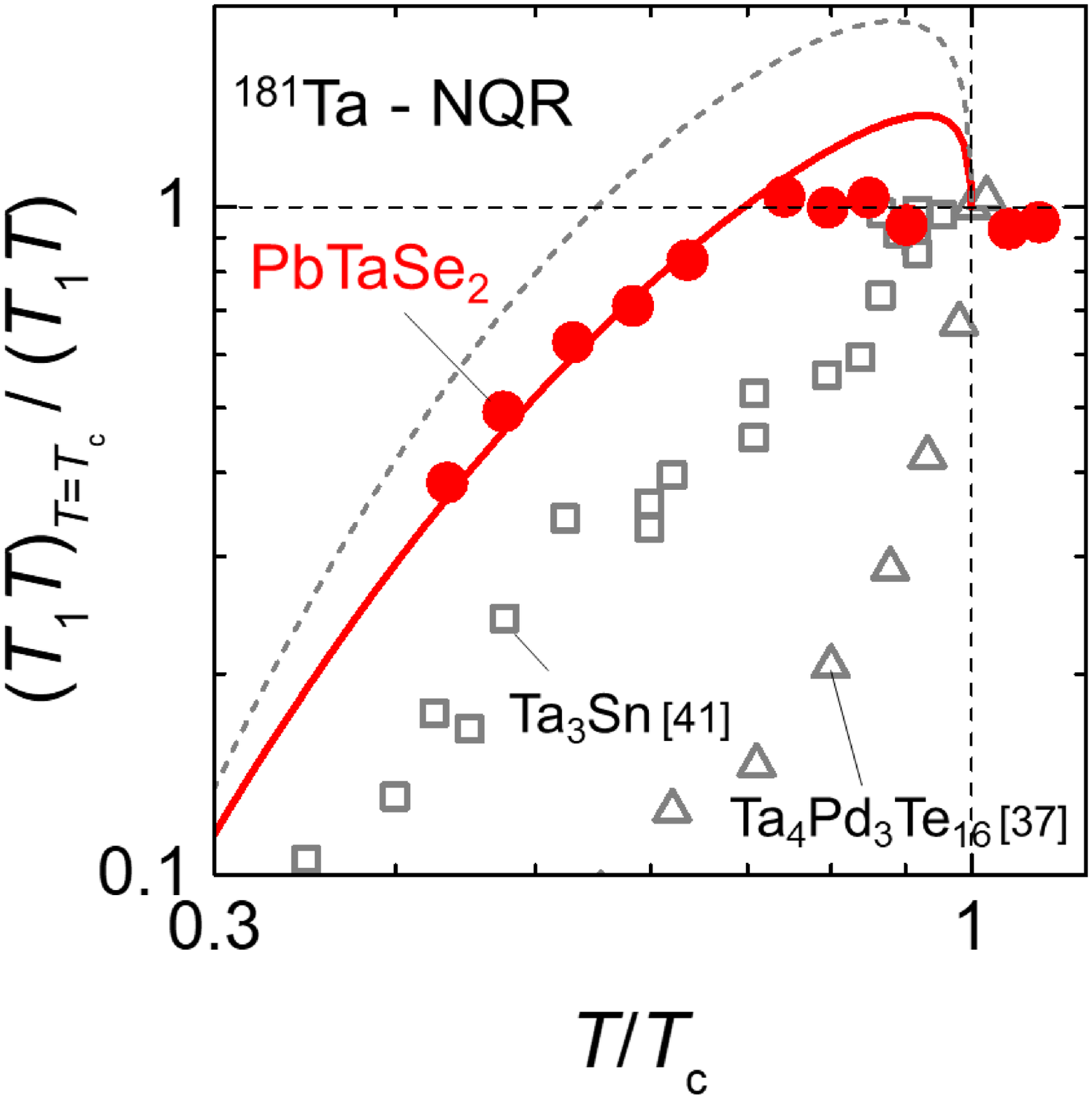} 
		\caption[Structure]{\label{iT1T}(Color online)
			$(T_{1}T)_{T=T_{\rm_c}}/(T_{1}T)$ {\it vs} $T/T_{\rm_c}$ probed by $^{181}$Ta-NQR for PbTaSe$_2$. Suppression of the coherence peak has been observed by $^{181}$Ta-NQR probes even in typical BCS superconductors like Ta$_3$Pd$_4$Te$_{16}$\cite{37} and Ta$_3$Sn\cite{41} due to the predominant nuclear quadrupole relaxation mechanism. Solid curve is the simulation of PbTaSe$_2$ using the BCS model, assuming the superposition of magnetic and quadrupole relaxation mechanisms in addition to the distribution of the superconducting gap (see in the text), which reproduces the experimental data well, in contrast to the simulation through only the magnetic relaxation mechanism (dashed curve).
}
		
	\end{center}
\end{figure}

In the superconducting state ($T < T_{\rm_c}$), $1/T_1$ shows exponential behavior with decreasing temperature, as seen in the Arrhenius plots of $T_1$ (see the inset of Fig. 3). 
We conclude that the superconducting state is dominated by the fully-gapped $s$-wave state in PbTaSe$_2$, which is consistent with previous results\cite{23,24,25,26,27,28,29,30,36}. 
The gap parameter $2\Delta$, estimated from the slope of the Arrhenius plot, is $\sim(3.09 \pm 0.07) k_{\rm_B}T_{\rm_c}$, which is slightly smaller than the value obtained in other experiments\cite{24,25,27,28,30}.

Here we comment on the smaller superconducting gap size in $^{181}$Ta-NQR at zero magnetic fields compared to $^{207}$Pb-NMR results\cite{30}, which is not attributed to the absence of vortex cores. 
It is worth mentioning that recent theoretical calculations suggested that pairing interaction varies among different FSs\cite{31}. 
This leads to the distribution of superconducting gap sizes in PbTaSe$_2$, which was previously discussed in terms of multiple or two gap structure in some experiments\cite{23,24,25,26,27,28,29,30}. 
According to their calculation, the gap size of the FSs dominated by Ta-5$d$ orbitals is relatively smaller than that of the FSs dominated by Pb-4$p$ orbitals\cite{31}.
Thus, the smaller gap size in the present Ta-NQR may indicate that the superconducting gaps in the FSs composed of dominant Ta-5$d$ orbitals were primarily probed assuming that the Ta-originated FSs may be strongly coupled with Ta nucleus. 
Further experiments are required to verify site-selective observation of specific superconducting gaps in multiband superconductors.

Finally, we discuss the temperature dependence of $1/T_1$. 
It is well known that even in BCS $s$-wave superconductors like Ta$_3$Pd$_4$Te$_{16}$\cite{37} and Ta$_3$Sn\cite{41}, $1/T_1$ obtained by $^{181}$Ta-NQR drops abruptly without coherence peak just below $T_{\rm_c}$, as shown in Fig. 4. 
This is because the quadrupole mechanism in nuclear spin relaxation process is dominant due to the large nuclear quadrupole momentum of the $^{181}$Ta nucleus. 
As shown in Fig. 4, the suppression of the coherence peak in PbTaSe$_2$ is not as remarkable in comparison with the other two Ta-based BCS superconductors, while prominent in comparison with the simulation curve of $s$-wave superconductivity with the gap size obtained by the Arrhenius plot (dashed curve in Fig. 4). 
This indicates that the quadrupole relaxation mechanism is minor in PbTaSe$_2$ although it affects the relaxation process to a certain degree. 
Furthermore, the temperature dependence of $1/T_1$ in PbTaSe$_2$ should be affected by the multiple gap structure mentioned above since distribution of superconducting gap size also results in the suppression of the coherence peak. 
Taking these points into account, we have fitted the temperature dependence of $1/T_1$ by changing the following parameters: the ratio of quadrupole relaxation mechanism ($R_{\rm_Q}$) to magnetic one ($R_{\rm_M}$), and the broadness of quasi-particle energies in the density of states ($\delta$)\cite{42}.
As shown in Figs. 3 and 4, the experimental results are reproduced well for $R_{\rm_Q}:R_{\rm_M}\sim2:8$ and $\delta/\Delta(0) \sim0.55$ (solid line in Figs. 3 and 4). 
The large broadening parameter $\delta/\Delta(0)$ is consistent with the essential broadening of gap energy proposed by the recent theoretical calculation that reveals the distribution of pairing interaction depending on each FS\cite{31}.

\section{Summary}
In summary, we performed a $^{181}$Ta-NQR measurement without magnetic fields in the non-centrosymmetric superconductor PbTaSe$_2$. 
Two sets of resonance peaks were observed in the $^{181}$Ta-NQR spectrum suggesting two inequivalent Ta sites. 
The DFT calculation identified the intrinsic Ta site in ideal $P$\={6}$m2$ structure, which enabled us to measure $T_1$ at the intrinsic site selectively. 
The exponential decrease in $1/T_1$ well below $T_{\rm_c}$ indicates a fully gapped superconducting state with superconducting gap size $2\Delta/(k_{\rm_B}T_{\rm_c}) = 3.09 \pm 0.07$. 
The smaller gap size obtained by $^{181}$Ta-NQR compared to $^{207}$Pb-NMR may suggest that the superconducting gaps in the Fermi surfaces attributed to Ta-5$d$ orbitals can be primarily probed, where the pairing interaction is calculated to be smaller than for FSs attributed to Pb-4$p$ orbitals. 
The temperature dependence of $1/T_1$ can be well explained by taking account of the gap size distribution proposed by theory\cite{31} in addition to the quadrupole relaxation mechanism. 
The {\it pure} $^{181}$Ta-NQR measurements in non-centrosymmetric PbTaSe$_2$ provide microscopic experimental evidence for the multiple superconducting gap properties inherent to multiband systems.

\section*{Aknowledgements}
We thank H. Usui for fruitful discussions. This work is partially supported by JSPS KAKENHI (Grant Nos. 16H04013, 18K18734, 16H06114, 18H04226, 16H06015 and 19H05173), JST PRESTO (No. JPMJPR16R2), the Murata Science Foundation, and the Mitsubishi Foundation.

\section*{$*$Corresponding author}
yokoi@gmr.phys.sci.osaka-u.ac.jp

murakawa@phys.sci.osaka-u.ac.jp

mukuda@mp.es.osaka-u.ac.jp

\end{document}